\documentclass[aps,prl,twocolumn,superscriptaddress]{revtex4}
\usepackage{amsmath,amssymb}
\usepackage{graphicx}
\usepackage{bm}

\begin{document}

\preprint{LA-UR-09-01607}

\def\CFA{CaFe$_2$As$_2$}
\def\musr{$\mu^+$SR}
\def\iTone{$T_1^{-1}$}

\title{NMR investigation of superconductivity and antiferromagnetism
in CaFe$_2$As$_2$ under pressure}

\author{S.-H. Baek}
\affiliation{Los Alamos National Laboratory, Los Alamos, NM 87545, USA}
\author{H. Lee}
\affiliation{Los Alamos National Laboratory, Los Alamos, NM 87545, USA}
\author{S. E. Brown}
\affiliation{Department of Physics, University of California, Los Angeles, CA 90095-1547, USA}
\author{N. J. Curro}
\affiliation{Department of Physics, University of California, Davis, CA 95616, USA}
\author{E. D. Bauer}
\affiliation{Los Alamos National Laboratory, Los Alamos, NM 87545, USA}
\author{F. Ronning}
\affiliation{Los Alamos National Laboratory, Los Alamos, NM 87545, USA}
\author{T. Park}
\affiliation{Los Alamos National Laboratory, Los Alamos, NM 87545, USA}
\affiliation{Department of Physics, Sungkyunkwan University, Suwon 440-746, Korea}
\author{J. D. Thompson}
\affiliation{Los Alamos National Laboratory, Los Alamos, NM 87545, USA}

\date{\today}

\begin{abstract}
We report $^{75}$As NMR measurements in CaFe$_2$As$_2$, made under applied
pressures up to 0.83 GPa produced by a standard clamp pressure cell. Our data reveal phase
segregation of paramagnetic (PM) and antiferromagnetic (AFM) phases over a range
of pressures, with the AFM phase more than 90 \% dominant at low temperatures.
{\it In situ} RF susceptibility measurements indicate the presence of
superconductivity.  $^{75}$As spin-lattice relaxation experiments indicate
that the $^{75}$As nuclei sample the superconductivity while in the 
magnetically-ordered environment. 
\end{abstract}

\pacs{76.60.-k, 74.10.+v, 74.62.Fj}

\maketitle

Like the cuprates, the newly-discovered FeAs pnictide family of
superconductors are notable for the emergence of a superconducting state with 
unusually high transition temperatures from a suppressed magnetic ground 
state. As such, it is important to establish the pairing interaction for each 
of the two systems, and 
further to identify what physical properties might be shared or different. For 
other systems in which superconducting ground states are stable only  
near to magnetically ordered ones, it is frequently suggested that magnetic
excitations in the (almost ordered) high symmetry phase are necessary for the
pairing, and more generally, the two order parameters may compete.
Signatures for the role of magnetic fluctuations could also manifest in
the order parameter symmetry. Consequently, the focus of many experimental studies is
to examine trends in properties in the vicinity that the magnetic
ordering temperature $T_N\to0$, which is accomplished through doping by
chemical substitution or through the application of high pressure.

The measurements reported here address the question of whether 
superconductivity exists in the antiferromagnetic (AFM) phase of the series of 
$A$Fe$_2$As$_2$ ($A$=Ca,Ba,Sr,Eu) compounds commonly 
referred to as {\it 122},
and are currently the subject of widespread activity, in part because
high quality single crystals are available. The onset of magnetic order
at a wavevector $\mathbf{Q}=(1,0,1)$ \cite{goldman08}
in the undoped compounds occurs discontinuously upon cooling, and coincides with a 
structural distortion lowering the lattice symmetry from tetragonal 
($\mathcal{T}$) 
to orthorhombic ($\mathcal{O}$). \CFA\ is notable because superconductivity 
was observed beyond relatively modest pressures, $\sim O$(0.5 GPa) 
\cite{torikachvili08,park08}. However, identification of the superconducting 
phase was complicated by the results of subsequent neutron scattering 
measurements \cite{kreyssig08}, which provided evidence for a 
transition to a so-called collapsed tetragonal ($c\mathcal{T}$) phase for
$P \gtrsim 0.4$ GPa. It was first presumed that superconductivity
occurs in the $c\mathcal{T}$ phase \cite{yildirim09,goko08}.
Later, it was shown that when He gas is used for
the pressure medium, a line of first order transitions separates the 
$\mathcal{O}$ and $c\mathcal{T}$ 
phases, but no superconductivity was found in either one under such conditions
\cite{yu09}. Evidently, the observation of superconductivity can only be
attributed to some (so far) unspecified non-hydrostatic conditions: in the neutron
scattering experiments \cite{goldman09}, as well as in \musr\ work
\cite{goko08}, there was evidence for
measurable volume fractions of segregated $\mathcal{O}$ and $c\mathcal{T}$
phases persisting over a wide range of applied pressures, 0.2--1.0~GPa.
Also, the sharp phase transition to the $\mathcal{O}$ phase seen at
ambient pressure broadened significantly when experiments were conducted in
standard clamp pressure cells \cite{canfield09,lee08}. The determination of
the superconducting phase remains an open question, and we note that 
phase segregation 
is not the only possibility; rather, coexistence is possible under some 
conditions in models where the SC state is competing for Fermi surface with 
the spin-density wave \cite{vorontsov09}.  

In the following, we present our results of $^{75}$As NMR measurements carried
out in a clamp cell. {\it In situ} rf susceptibility measurements indicate the 
presence of superconductivity with significant shielding fractions. 
NMR spectra recorded at low temperatures indicate the majority volume 
fraction is in the AFM 
$\mathcal{O}$ phase. Some details of the observations are inconsistent with 
the proposal for coexistence. In particular, no significant moment reduction in 
the $\mathcal{O}$ phase is observed relative to the ambient pressure case, and 
the 
wavevector is commensurate and unchanged over the entire range of pressures
studied. Further, while superconductivity is clearly detectable in the
spin-lattice relaxation data for nuclei situated in the $\mathcal{O}$ phase,
its origin is likely in vortex dynamics rather than hyperfine fields. We
conclude that the detectable nuclei are at least close to a superconducting
volume within distances of order of the penetration depth.

Single crystals of \CFA\ were prepared as described in Ref.~\cite{ronning08}.
$^{75}$As ($I=3/2$) nuclear magnetic resonance (NMR) measurements were
performed in a standard BeCu clamp-type pressure cell, using a silicone fluid
for the medium. The pressures reported were determined by measuring
the $^{63}$Cu NQR frequency in Cu$_2$O powder sample at helium temperature
\cite{reyes92}; the pressure set at $T=300$ K is approximately 0.15-0.20 GPa or
higher for all cases. For all measurements described here, the external field is applied
along the $c$-axis of the sample.

Fig.~1(a) shows the temperature evolution of the central transition
$^{75}$As spectrum at $P=0.34$ GPa at the fixed frequency $\nu=42$ MHz. The
spectral intensities $I$ are multiplied by temperature for the purpose of
relative comparison. For $T<135$ K, the contribution to $IT$ from the
paramagnetic (PM) $\mathcal{T}$ phase starts to decrease but becomes negligibly small
only at much lower temperatures. This behavior contrasts with the complete
disappearance of the paramagnetic PM component below $T_N$ at ambient pressure
\cite{baek09}. Coincident with the loss of PM signal, we observe the
appearance and increase of signal intensity characteristic of the
antiferromagnetically-ordered (AFM) $\mathcal{O}$ phase. The temperature evolution of
$I(P=0.34 \text{ GPa},T)T$ across the spectrum, normalized at $T=140$ K, is shown in
the inset of Fig.~1(a). The trend in the spectra is clearly accounted for by a
decreasing volume fraction of $\mathcal{T}$-phase, accompanied by an associated increase
in $\mathcal{O}$ phase. Similar results were obtained for $P=0.5$ GPa. At the same time,
the internal field at the $^{75}$As nuclei in the $\mathcal{O}$ phase is nearly
independent of pressure, as shown in Fig.~1(b). As this quantity is the
product of the hyperfine coupling constant and thermally averaged Fe spin
moment, the observation indicates that the magnetic moment of $\mathcal{O}$-phase
Fe is robust against applied pressure.

\begin{figure}
\label{fig:1}
\centering
\includegraphics[width=\linewidth]{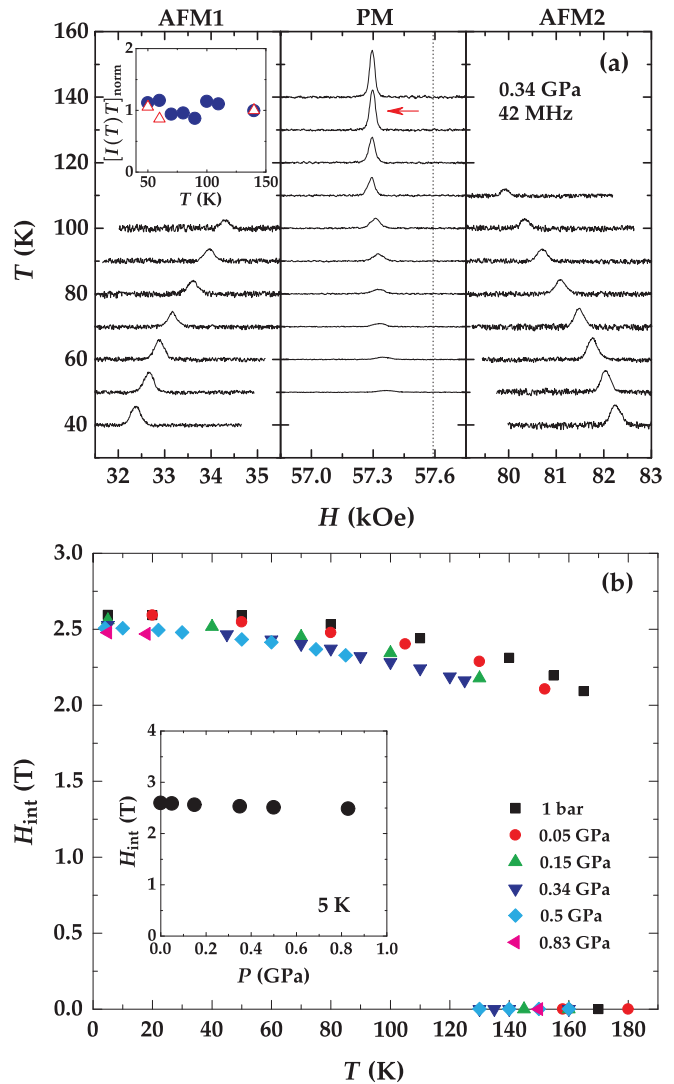}
\caption{(a) Field-swept $^{75}$As spectra vs $T$
at $P=0.34$ GPa. In all cases, the carrier frequency $\nu=42$ MHz. The
intensity characteristic of the paramagnetic (PM) line appears in the center panel.
Spectral intensity appearing in the side panels are shifted by hyperfine
fields produced by the AFM ordering, and scaled $\times 3$ for clarity. The
phase change, from PM$\to$AFM, occurs over a wide range of temperatures. The
inset shows that the total nuclear magnetization from the three contributions,
follows a Curie Law to within experimental uncertainty. Triangles are from 
data at $P=0.5$ GPa. (b) Internal field
$H_\text{int}$ obtained from the magnetic field separation between AFM and PM
absorption locations. The inset shows the pressure dependence of $H_\text{int}$
at $T=5$ K.}
\end{figure}

In Fig.~2(a) is a summary of the effect of $P$ and $T$ on the PM fraction. The
range of temperatures over which two phases are observed broadens with
increasing $P$ \cite{footnote1}. For $P>0.3$ GPa, we take note that the
decrease in PM volume fraction is relatively fast initially below $T_N$, then
slows below 100 K. Accompanying the slower dropoff is a considerable
broadening of the PM signal. $T^*$ is defined as the temperature below which the PM
fraction falls below 10\% of the full intensity. Measurements of the PM component are reported only for the case that the measurement bandwidth exceeds the linewidth.

The rapid decrease of PM volume just below $T_N$ is due to a nonuniform, progressive
transformation to $\mathcal{O}$ phase.  The 
broadening of PM line in Fig.~1 (a) and the long tail in Fig.~2 (a) seen 
below 100 K is tentatively ascribed to the presence of the remaining small 
volume fraction of $c\mathcal{T}$ 
phase. This means that, at $P=0.83$ GPa, the volume of $c\mathcal{T}$ phase is largest
at intermediate temperature and extrapolates to a negligible fraction
for low $T\to0$. Thus, the magnetic $\mathcal{O}$ phase is still dominant up to 0.83 GPa at low
$T$, even though independent transport measurements showed the clear signature of the
$\mathcal{T}-c\mathcal{T}$ transition
at 150 K at this pressure \cite{lee08}. The slow decrease of PM fraction and the associated
line broadening for $T<100$ K is consistent with recent neutron scattering
results (see Fig.~4 in Ref.~\cite{goldman09}).

\begin{figure}
\label{fig:2}
\centering
\includegraphics[width=\linewidth]{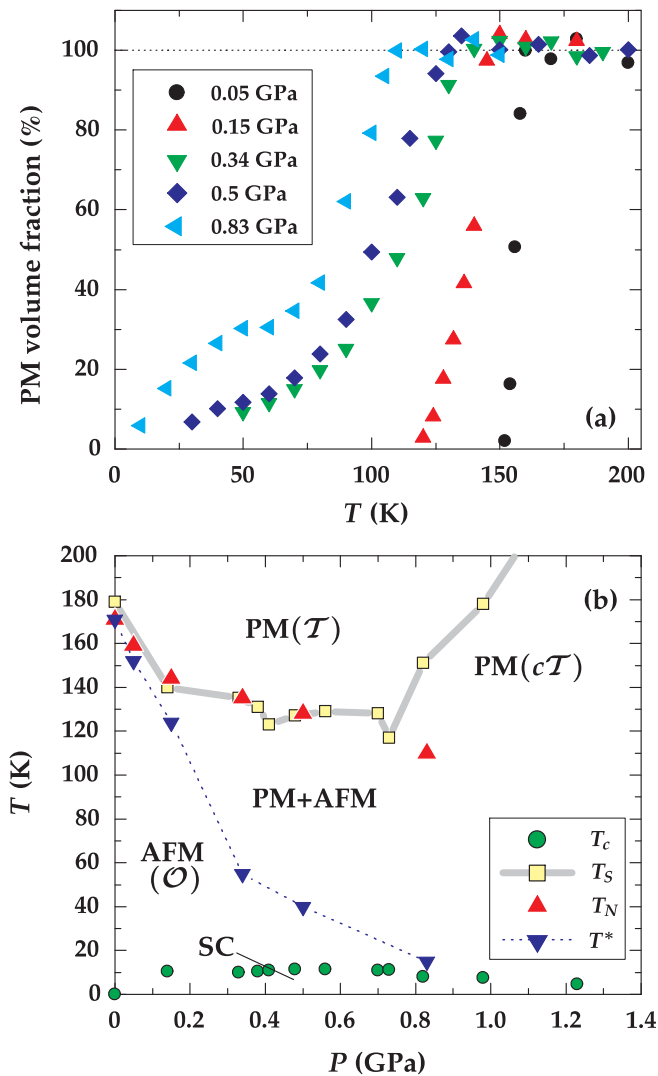}
\caption{(a) PM volume fraction versus temperature as a function of pressure,
obtained by monitoring the temperature variation of the PM intensity
multiplied by $T$. (b) Phase diagram obtained from (a). $T^*$ represents the
temperature at which PM intensity is 10 \% of the full intensity. Also shown
is a line of superconducting transitions at $T_c$, obtained from transport
experiments made under similar clamp-cell conditions \cite{lee08}.}
\end{figure}

The $P-T$ phase diagram obtained by monitoring the PM intensity is shown in
Fig. 2(b). Included also are the structural transition temperature $T_S$ and
the SC transition temperature $T_c$, inferred independently from resistivity
data \cite{lee08}. As marked in the figure, $>90$\% volume fraction is accounted for in the
magnetic $\mathcal{O}$ phase at the superconducting transition temperature at
all pressures investigated. It would seem that the possibility that the SC is
occurring in the minority PM volume in filamentary/grain boundary form is unlikely since the
response of the NMR tank circuit, and the spin-lattice relaxation rate of the
$^{75}$As spins situated in the magnetic phase, both show clear signatures for
the presence of superconductivity. Below we describe the results of these
measurements.

\begin{figure}
\label{fig:3}
\centering
\includegraphics[width=\linewidth]{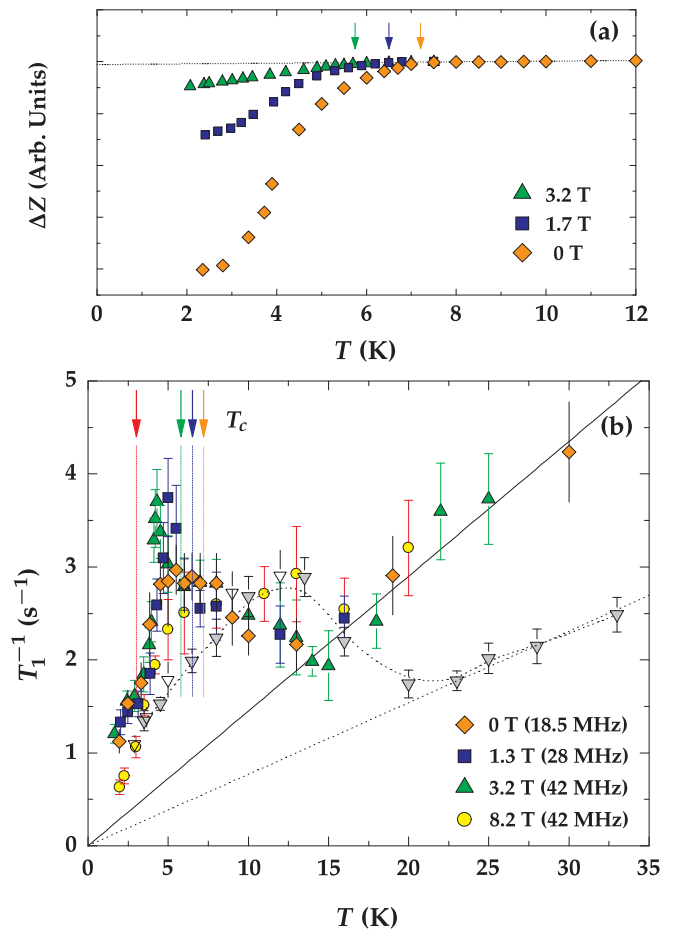}
\caption{(a) Loss of impedance $\Delta Z$ due to superconductivity measured in 
NMR tank circuit at 0.5 GPa. Down arrows denote $T_c$ for each 
field.  (b) $^{75}$As $T_1^{-1}$ versus $T$ at $P=0.5$ GPa. Down triangles are
data taken at ambient pressure at 45 MHz (filled: 3.6 T, open: 8.8 T).}
\end{figure}
The rf response of the NMR tank circuit confirming the presence of
superconductivity is shown in Fig. 3 (a).  The zero-field SC
onset at $T_c=7.2$ K is in agreement with the transport measurements, as is
the effect of an applied field ($dT_c/dH=0.5$ K/T) \cite{park08}. The spin 
lattice relaxation, recorded for the two AFM lines [Fig.~1(a)], exhibits 
features correlated with $T_c$. $T_1^{-1}$ ($P=0.5$ GPa) 
at the low-field line (AFM1, $B= 1.3 \text{ and } 3.2$ T), shown in Fig.~3 (b),
is described by
$(T_1T)^{-1}=\text{constant}$ at high temperature, but is enhanced relative to
this form for $T<15$ K. Also visible is a sharp peak superimposed on the broad
feature and confined to a narrow temperature range just below $T_c$. In an
attempt to associate directly the observation with superconductivity, the same
measurement was carried out at $B=8.2$ T on the AFM2 line while maintaining
$\nu=42$ MHz, and in zero field at 18.5 MHz.
The larger applied field completely eliminates all evidence for
the sharp peak, although the broad enhancement remains. Similarly, the sharp maximum is absent from the zero field experiments. Also shown in Fig.~3 (b)
are results from ambient pressure experiments for which there is no
superconductivity; a similar broad enhancement of $T_1^{-1}$ is still
visible.

The variation of the relaxation rate at low temperatures, and sampling
different fields and pressures, indicates that the sharp maximum in \iTone\ is 
linked to the superconducting transition, but likely results from vortex 
dynamics since it is absent at both high field and zero field. The broad 
enhancement of $T_1^{-1}$ is present under all conditions, whether the sample is 
superconducting or not: 
evidently the broad peak is not a sufficient predictor for superconductivity. We
do not comment further, except to note that the origin of the broad relaxation
enhancement appears to be magnetic, possibly reflecting glassy dynamics of
antiferromagnetic domains \cite{curro09}, and it is reminiscent of observations
made in other FeAs superconductors \cite{nakai08,imai09}.
A key question is whether the hyperfine
fields in the $\mathcal{O}$ phase are influenced by the onset of the SC state.
This onset could be manifested in terms of a spectral shift, a change in the
amplitude or wavevector of the magnetic ordering, or in the hyperfine
contribution to the relaxation rate. Presumably, the contributions to the
relaxation rate are tuned by the magnetic field. In fact, no changes 
attributed to the onset of superconductivity {\it and} originating 
specifically with the hyperfine fields are 
observed. Therefore, we conclude that the
$\mathcal{O}$ phase is not superconducting in isolation, even though here it 
is the (overwhelming) 
majority phase, and the observed shielding fraction is significant.
On the other hand, the Korringa relaxation (above $\sim 15$ K) at 0.5 GPa is
twice that at atmospheric pressure, implying a factor of $\sqrt{2}$ larger density of
states $N(0)\propto m_\text{eff} n^{1/3}$, where $n$ is the carrier density
and $m_\text{eff}$ is their effective mass.   Whereas, the broad enhancement of
$T_1^{-1}$ below $\sim 15$ K is not a sufficient condition for
superconductivity, the pressure-induced increase of $N(0)$ \textit{in the
orthorhombic phase} may be a necessary requirement.

Finally, we comment on the observation of majority $\mathcal{O}$ phase
at the pressures investigated in this study. Similar volume fractions are
evident in neutron scattering studies \cite{goldman09}, and these were
attributed to non-hydrostatic conditions. To determine more specifically what
aspects stabilize the $\mathcal{O}$ phase over the $c\mathcal{T}$ phase will require
producing the relevant stress(es) under controlled conditions. For example, 
the structural order parameter $\delta=(a-b)/(a+b)$ will couple to an in-plane 
stress difference $\sigma_{xx}-\sigma_{yy}$. Consequently, the $\mathcal{O}$ 
phase is stabilized by this coupling, and particularly so when the transition 
is weakly first order. Variable stresses can lead to a broadening of the 
structural transition, and possibly to the predominance of the $\mathcal{O}$ 
phase at low temperatures and intermediate pressures. Similar reasoning could 
lead to small amounts of $\mathcal{T}$ phase at low temperatures, even though 
it is not a ground state under hydrostatic conditions. 
Thus, we cannot 
distinguish between two possibilities for superconductivity: it could occur in 
a small fraction of $\mathcal{T}$ and/or $c\mathcal{T}$ phase(s), or in the 
majority $\mathcal{O}$ phase.  
With respect to the latter possibility, we note 
that in ref.~\cite{lee08}, the detection of superconductivity in the clamped
cells is attributed to doping into the $\mathcal{O}$ phase from a minority $c\mathcal{T}$
phase, and the increase of $(T_1T)^{-1}$ by approximately factor two with
the application of pressure could be associated with such a change in carrier
concentration.

We thank Hironori Sakai for useful discussions. S.E.B is grateful for the 
hospitality and support of the Condensed Matter and Thermal Physics group at Los Alamos 
National Laboratory. Work at Los Alamos National 
Laboratory was performed under the auspices of the US Department of Energy,
Office of Science. Work at UCLA was supported in part by the National Science
Foundation under grant number DMR-0804625.

\bibliography{/mydocuments/mydata/mybib}

\begin{thebibliography}{17}
\expandafter\ifx\csname natexlab\endcsname\relax\def\natexlab#1{#1}\fi
\expandafter\ifx\csname bibnamefont\endcsname\relax
  \def\bibnamefont#1{#1}\fi
\expandafter\ifx\csname bibfnamefont\endcsname\relax
  \def\bibfnamefont#1{#1}\fi
\expandafter\ifx\csname citenamefont\endcsname\relax
  \def\citenamefont#1{#1}\fi
\expandafter\ifx\csname url\endcsname\relax
  \def\url#1{\texttt{#1}}\fi
\expandafter\ifx\csname urlprefix\endcsname\relax\def\urlprefix{URL }\fi
\providecommand{\bibinfo}[2]{#2}
\providecommand{\eprint}[2][]{\url{#2}}

\bibitem[{\citenamefont{Goldman et~al.}(2008)\citenamefont{Goldman, Argyriou,
  Ouladdiaf, Chatterji, Kreyssig, Nandi, Ni, Bud'ko, Canfield, and
  McQueeney}}]{goldman08}
\bibinfo{author}{\bibfnamefont{A.~I.} \bibnamefont{Goldman}},
  \bibinfo{author}{\bibfnamefont{D.~N.} \bibnamefont{Argyriou}},
  \bibinfo{author}{\bibfnamefont{B.}~\bibnamefont{Ouladdiaf}},
  \bibinfo{author}{\bibfnamefont{T.}~\bibnamefont{Chatterji}},
  \bibinfo{author}{\bibfnamefont{A.}~\bibnamefont{Kreyssig}},
  \bibinfo{author}{\bibfnamefont{S.}~\bibnamefont{Nandi}},
  \bibinfo{author}{\bibfnamefont{N.}~\bibnamefont{Ni}},
  \bibinfo{author}{\bibfnamefont{S.~L.} \bibnamefont{Bud'ko}},
  \bibinfo{author}{\bibfnamefont{P.~C.} \bibnamefont{Canfield}},
  \bibnamefont{and} \bibinfo{author}{\bibfnamefont{R.~J.}
  \bibnamefont{McQueeney}}, \bibinfo{journal}{Phys. Rev. B}
  \textbf{\bibinfo{volume}{78}}, \bibinfo{pages}{100506}
  (\bibinfo{year}{2008}).

\bibitem[{\citenamefont{Torikachvili et~al.}(2008)\citenamefont{Torikachvili,
  Bud'ko, Ni, and Canfield}}]{torikachvili08}
\bibinfo{author}{\bibfnamefont{M.~S.} \bibnamefont{Torikachvili}},
  \bibinfo{author}{\bibfnamefont{S.~L.} \bibnamefont{Bud'ko}},
  \bibinfo{author}{\bibfnamefont{N.}~\bibnamefont{Ni}}, \bibnamefont{and}
  \bibinfo{author}{\bibfnamefont{P.~C.} \bibnamefont{Canfield}},
  \bibinfo{journal}{Phys. Rev. Lett.} \textbf{\bibinfo{volume}{101}},
  \bibinfo{pages}{057006} (\bibinfo{year}{2008}).

\bibitem[{\citenamefont{Park et~al.}(2008)\citenamefont{Park, Park, Lee,
  Klimczuk, Bauer, Ronning, and Thompson}}]{park08}
\bibinfo{author}{\bibfnamefont{T.}~\bibnamefont{Park}},
  \bibinfo{author}{\bibfnamefont{E.}~\bibnamefont{Park}},
  \bibinfo{author}{\bibfnamefont{H.}~\bibnamefont{Lee}},
  \bibinfo{author}{\bibfnamefont{T.}~\bibnamefont{Klimczuk}},
  \bibinfo{author}{\bibfnamefont{E.~D.} \bibnamefont{Bauer}},
  \bibinfo{author}{\bibfnamefont{F.}~\bibnamefont{Ronning}}, \bibnamefont{and}
  \bibinfo{author}{\bibfnamefont{J.~D.} \bibnamefont{Thompson}},
  \bibinfo{journal}{J. Phys. Condens. Matter} \textbf{\bibinfo{volume}{20}},
  \bibinfo{pages}{322204} (\bibinfo{year}{2008}).

\bibitem[{\citenamefont{Kreyssig et~al.}(2008)\citenamefont{Kreyssig, Green,
  Lee, Samolyuk, Zajdel, Lynn, Bud'ko, Torikachvili, Ni, Nandi
  et~al.}}]{kreyssig08}
\bibinfo{author}{\bibfnamefont{A.}~\bibnamefont{Kreyssig}},
  \bibinfo{author}{\bibfnamefont{M.~A.} \bibnamefont{Green}},
  \bibinfo{author}{\bibfnamefont{Y.}~\bibnamefont{Lee}},
  \bibinfo{author}{\bibfnamefont{G.~D.} \bibnamefont{Samolyuk}},
  \bibinfo{author}{\bibfnamefont{P.}~\bibnamefont{Zajdel}},
  \bibinfo{author}{\bibfnamefont{J.~W.} \bibnamefont{Lynn}},
  \bibinfo{author}{\bibfnamefont{S.~L.} \bibnamefont{Bud'ko}},
  \bibinfo{author}{\bibfnamefont{M.~S.} \bibnamefont{Torikachvili}},
  \bibinfo{author}{\bibfnamefont{N.}~\bibnamefont{Ni}},
  \bibinfo{author}{\bibfnamefont{S.}~\bibnamefont{Nandi}},
  \bibnamefont{et~al.}, \bibinfo{journal}{Phys. Rev. B}
  \textbf{\bibinfo{volume}{78}}, \bibinfo{pages}{184517}
  (\bibinfo{year}{2008}).

\bibitem[{\citenamefont{Yildirim}(2009)}]{yildirim09}
\bibinfo{author}{\bibfnamefont{T.}~\bibnamefont{Yildirim}},
  \bibinfo{journal}{Phys. Rev. Lett.} \textbf{\bibinfo{volume}{102}},
  \bibinfo{pages}{037003} (\bibinfo{year}{2009}).

\bibitem[{\citenamefont{Goko et~al.}()\citenamefont{Goko, Aczel,
  Baggio-Saitovitch, Bud'ko, Canfield, Carlo, Chen, Dai, Hamann, Hu
  et~al.}}]{goko08}
\bibinfo{author}{\bibfnamefont{T.}~\bibnamefont{Goko}},
  \bibinfo{author}{\bibfnamefont{A.~A.} \bibnamefont{Aczel}},
  \bibinfo{author}{\bibfnamefont{E.}~\bibnamefont{Baggio-Saitovitch}},
  \bibinfo{author}{\bibfnamefont{S.~L.} \bibnamefont{Bud'ko}},
  \bibinfo{author}{\bibfnamefont{P.~C.} \bibnamefont{Canfield}},
  \bibinfo{author}{\bibfnamefont{J.~P.} \bibnamefont{Carlo}},
  \bibinfo{author}{\bibfnamefont{G.~F.} \bibnamefont{Chen}},
  \bibinfo{author}{\bibfnamefont{P.}~\bibnamefont{Dai}},
  \bibinfo{author}{\bibfnamefont{A.~C.} \bibnamefont{Hamann}},
  \bibinfo{author}{\bibfnamefont{W.~Z.} \bibnamefont{Hu}},
  \bibnamefont{et~al.}, \eprint{arXiv:0808.1425 (unpublished)}.

\bibitem[{\citenamefont{Yu et~al.}(2009)\citenamefont{Yu, Aczel, Williams,
  Bud'ko, Ni, Canfield, and Luke}}]{yu09}
\bibinfo{author}{\bibfnamefont{W.}~\bibnamefont{Yu}},
  \bibinfo{author}{\bibfnamefont{A.~A.} \bibnamefont{Aczel}},
  \bibinfo{author}{\bibfnamefont{T.~J.} \bibnamefont{Williams}},
  \bibinfo{author}{\bibfnamefont{S.~L.} \bibnamefont{Bud'ko}},
  \bibinfo{author}{\bibfnamefont{N.}~\bibnamefont{Ni}},
  \bibinfo{author}{\bibfnamefont{P.~C.} \bibnamefont{Canfield}},
  \bibnamefont{and} \bibinfo{author}{\bibfnamefont{G.~M.} \bibnamefont{Luke}},
  \bibinfo{journal}{Phys. Rev. B} \textbf{\bibinfo{volume}{79}},
  \bibinfo{pages}{020511} (\bibinfo{year}{2009}).

\bibitem[{\citenamefont{Goldman et~al.}(2009)\citenamefont{Goldman, Kreyssig,
  Proke\v{s}, Pratt, Argyriou, Lynn, Nandi, Kimber, Chen, Lee
  et~al.}}]{goldman09}
\bibinfo{author}{\bibfnamefont{A.~I.} \bibnamefont{Goldman}},
  \bibinfo{author}{\bibfnamefont{A.}~\bibnamefont{Kreyssig}},
  \bibinfo{author}{\bibfnamefont{K.}~\bibnamefont{Proke\v{s}}},
  \bibinfo{author}{\bibfnamefont{D.~K.} \bibnamefont{Pratt}},
  \bibinfo{author}{\bibfnamefont{D.~N.} \bibnamefont{Argyriou}},
  \bibinfo{author}{\bibfnamefont{J.~W.} \bibnamefont{Lynn}},
  \bibinfo{author}{\bibfnamefont{S.}~\bibnamefont{Nandi}},
  \bibinfo{author}{\bibfnamefont{S.~A.~J.} \bibnamefont{Kimber}},
  \bibinfo{author}{\bibfnamefont{Y.}~\bibnamefont{Chen}},
  \bibinfo{author}{\bibfnamefont{Y.~B.} \bibnamefont{Lee}},
  \bibnamefont{et~al.}, \bibinfo{journal}{Phys. Rev. B}
  \textbf{\bibinfo{volume}{79}}, \bibinfo{pages}{024513}
  (\bibinfo{year}{2009}).

\bibitem[{\citenamefont{Canfield et~al.}()}]{canfield09}
\bibinfo{author}{\bibfnamefont{P.~C.} \bibnamefont{Canfield}}
  \bibnamefont{et~al.}, \eprint{arXiv:0901.4672 (unpublished)}.

\bibitem[{\citenamefont{Vorontsov et~al.}(2009)\citenamefont{Vorontsov,
  Vavilov, and Chubukov}}]{vorontsov09}
\bibinfo{author}{\bibfnamefont{A.~B.} \bibnamefont{Vorontsov}},
  \bibinfo{author}{\bibfnamefont{M.~G.} \bibnamefont{Vavilov}},
  \bibnamefont{and} \bibinfo{author}{\bibfnamefont{A.~V.}
  \bibnamefont{Chubukov}}, \bibinfo{journal}{Phys. Rev. B}
  \textbf{\bibinfo{volume}{79}}, \bibinfo{pages}{060508}
  (\bibinfo{year}{2009}).

\bibitem[{\citenamefont{Ronning et~al.}(2008)\citenamefont{Ronning, Klimczuk,
  Bauer, Volz, and Thompson}}]{ronning08}
\bibinfo{author}{\bibfnamefont{F.}~\bibnamefont{Ronning}},
  \bibinfo{author}{\bibfnamefont{T.}~\bibnamefont{Klimczuk}},
  \bibinfo{author}{\bibfnamefont{E.~D.} \bibnamefont{Bauer}},
  \bibinfo{author}{\bibfnamefont{H.}~\bibnamefont{Volz}}, \bibnamefont{and}
  \bibinfo{author}{\bibfnamefont{J.~D.} \bibnamefont{Thompson}},
  \bibinfo{journal}{J. Phys. Condens. Matter} \textbf{\bibinfo{volume}{20}},
  \bibinfo{pages}{322201} (\bibinfo{year}{2008}).

\bibitem[{\citenamefont{Reyes et~al.}(1992)\citenamefont{Reyes, Ahrens,
  Heffner, Hammel, and Thompson}}]{reyes92}
\bibinfo{author}{\bibfnamefont{A.~P.} \bibnamefont{Reyes}},
  \bibinfo{author}{\bibfnamefont{E.~T.} \bibnamefont{Ahrens}},
  \bibinfo{author}{\bibfnamefont{R.~H.} \bibnamefont{Heffner}},
  \bibinfo{author}{\bibfnamefont{P.~C.} \bibnamefont{Hammel}},
  \bibnamefont{and} \bibinfo{author}{\bibfnamefont{J.~D.}
  \bibnamefont{Thompson}}, \bibinfo{journal}{Rev. Sci. Intrum.}
  \textbf{\bibinfo{volume}{63}}, \bibinfo{pages}{3120} (\bibinfo{year}{1992}).

\bibitem[{\citenamefont{Baek et~al.}(2009)\citenamefont{Baek, Curro, Klimczuk,
  Bauer, Ronning, and Thompson}}]{baek09}
\bibinfo{author}{\bibfnamefont{S.-H.} \bibnamefont{Baek}},
  \bibinfo{author}{\bibfnamefont{N.~J.} \bibnamefont{Curro}},
  \bibinfo{author}{\bibfnamefont{T.}~\bibnamefont{Klimczuk}},
  \bibinfo{author}{\bibfnamefont{E.~D.} \bibnamefont{Bauer}},
  \bibinfo{author}{\bibfnamefont{F.}~\bibnamefont{Ronning}}, \bibnamefont{and}
  \bibinfo{author}{\bibfnamefont{J.~D.} \bibnamefont{Thompson}},
  \bibinfo{journal}{Phys. Rev. B} \textbf{\bibinfo{volume}{79}},
  \bibinfo{pages}{052504} (\bibinfo{year}{2009}).

\bibitem[{foo()}]{footnote1}
\bibinfo{note}{The inhomogeneous coexistence of AFM and PM may persist even at
  ambient pressure in 1--2 K range, and would explain the continuous magnetic
  transition in neutron diffraction \cite{goldman08}, despite the clear
  first-order transition.}

\bibitem[{\citenamefont{Curro et~al.}()}]{curro09}
\bibinfo{author}{\bibfnamefont{N.~J.} \bibnamefont{Curro}}
  \bibnamefont{et~al.}, \eprint{arXiv:0902.4492 (unpublished)}.

\bibitem[{\citenamefont{Nakai et~al.}(2008)\citenamefont{Nakai, Ishida,
  Kamihara, Hirano, and Hosono}}]{nakai08}
\bibinfo{author}{\bibfnamefont{Y.}~\bibnamefont{Nakai}},
  \bibinfo{author}{\bibfnamefont{K.}~\bibnamefont{Ishida}},
  \bibinfo{author}{\bibfnamefont{Y.}~\bibnamefont{Kamihara}},
  \bibinfo{author}{\bibfnamefont{M.}~\bibnamefont{Hirano}}, \bibnamefont{and}
  \bibinfo{author}{\bibfnamefont{H.}~\bibnamefont{Hosono}},
  \bibinfo{journal}{Phys. Rev. Lett.} \textbf{\bibinfo{volume}{101}},
  \bibinfo{pages}{077006} (\bibinfo{year}{2008}).

\bibitem[{\citenamefont{Imai et~al.}(2009)\citenamefont{Imai, Ahilan, Ning,
  McQueen, and Cava}}]{imai09}
\bibinfo{author}{\bibfnamefont{T.}~\bibnamefont{Imai}},
  \bibinfo{author}{\bibfnamefont{K.}~\bibnamefont{Ahilan}},
  \bibinfo{author}{\bibfnamefont{F.~L.} \bibnamefont{Ning}},
  \bibinfo{author}{\bibfnamefont{T.~M.} \bibnamefont{McQueen}},
  \bibnamefont{and} \bibinfo{author}{\bibfnamefont{R.~J.} \bibnamefont{Cava}},
  \bibinfo{journal}{Phys. Rev. Lett.} \textbf{\bibinfo{volume}{102}},
  \bibinfo{pages}{177005} (\bibinfo{year}{2009}).

\end{thebibliography}
\end{document}